\documentclass[12pt]{iopart}

\usepackage{epsfig}  
\begin{document}

\title{Understanding the tsunami with a simple model}

\author{O Helene}

\address{Instituto de F\'\i sica da Universidade de S\~ao Paulo, C.P. 66318, CEP 05315-970, 
S\~ao Paulo, Brazil}

\author{M T Yamashita}

\address{Universidade Estadual Paulista, CEP 18409-010, Itapeva/SP, Brazil}

\begin{abstract}
In this paper, we use the approximation of shallow water waves (Margaritondo 
2005 {\it Eur. J. Phys.} {\bf 26} 401) to understand the behavior of a tsunami in a variable 
depth. We deduce the shallow water wave equation and the continuity equation that must be 
satisfied when a wave encounters a discontinuity in the sea depth. A short explanation about how 
the tsunami hit the west coast of India is given based on the refraction phenomenon. Our procedure 
also includes a simple numerical calculation suitable for undergraduate students in physics and engineering.
\end{abstract}

\maketitle

\section{Introduction}

Tsunamis are water waves with long wavelengths that can be triggered by submarine earthquakes, 
landslides, volcanic eruption and large asteroid impacts. These non-dispersive waves can travel 
for thousands of kilometers from the disturbance area where they have been created with a minimum loss of energy. 
As any wave, tsunamis can be reflected, transmitted, refracted and diffracted.

The physics of a tsunami can be very complex, especially if we consider its creation and behavior next to the 
beach, where it can break. However, since tsunamis are composed of waves with very large wavelengths, sometimes 
greater than 100 km, they can be considered as shallow waves, even in oceans with depths of a few kilometers.
 
The shallow water approximation simplifies considerably the problem and still allows us to understand a lot of 
the physics of a tsunami. Using such approximation, Margaritondo \cite{MaEJP05} deduced the dispersion relation 
of tsunami waves extending a model developed by Behroozi and Podolefsky \cite{BeEJP01}. Since energy losses due 
to viscosity and friction at the bottom \cite{CrAJP87} can be neglected in the case of shallow waves, Margaritondo, 
considering energy conservation, explained the increase of the wave height when a tsunami approaches the coast, 
where the depth of the sea and the wave velocity are both reduced.

In this paper we use one of the results of Ref. \cite{MaEJP05} in order to deduce the wave equation and include the 
variation of the seabed. Thus, we are able to explain the increase of the wave amplitude when passing from deeper 
to shallow water. Also, we discuss the refraction of tsunami waves. This phenomenon allowed the tsunami of 
December 24, 2004, created at the Bay of Bengal, to hit the west coast of India (a detailed description is given 
in \cite{LaSc05}). These both inclusions - the seabed topography and the wave refraction - where pointed out by 
Chu \cite{chu} as necessary to understand some other phenomena observed in tsunamis.

This paper is organized as follows. The wave equation and the water flux conservation are used in section 2 in order 
to explain how and how much a shallow wave increases when passing from a deeper to a shallow water. In section 3, 
we extend the results obtained in section 2 to study how a wave packet propagates in a water tank where the depth
varies; in this section we use some numerical procedures that can be extended to the study of any wave propagating 
in a non-homogeneous medium. Also, the refraction of the 2004 tsunami in the south on India is discussed in section 3. The 
shallow wave and the continuity equations are deduced in appendix A.

\section{Reflection and transmission of waves in one dimension}

Consider a perturbation on the water surface in a rectangular tank with a constant 
depth. In the limit of large wavelengths and a small amplitude compared with the depth 
of the tank, the wave equation can be simplified to (see Appendix A)
\begin{equation}
\frac{\partial^2y}{\partial t^2}=gh\frac{\partial^2y}{\partial x^2},
\label{wave}
\end{equation}
where $y(x,t)$ is the vertical displacement of the water surface at a time $t$, 
propagating in the $x$ direction, $g$ is the gravity acceleration and $h$ is the water depth.

Equation (\ref{wave}) is the most common one-dimensional wave equation. It is a second-order 
linear partial differential equation and, since $g$ and $h$ are constants, any function 
$y=f(x\pm vt)$ is a solution ($v=\sqrt{gh}$ is the wave velocity). An interesting aspect 
of eq. (\ref{wave}) is that a propagating pulse does not present a dispersion due to the same 
velocity of all wavelengths and, thus, preserves its shape. Light in vacuum (and, in a very 
good approximation, in air) is non-dispersive. Also, sound waves in the air are nearly non-dispersive. 
(If dispersion was important in the propagation of the sound in the air, a sound would be heard 
different in different positions, i.e., music and conversation would be impossible)

However, the velocity of shallow-water wave varies with the depth. Thus, shallow-water waves are
dispersive in a non-uniform seadepth.

In order to study the evolution of a tsunami in a rectangular box with variable depth, which will be 
detailed in the next section, we approximate the irregular depth by successive steps. So, in the next 
paragraphs we will explain the treatment used when a wave encounters a discontinuity.

\begin{figure}[ht]
\centerline{\epsfig{figure=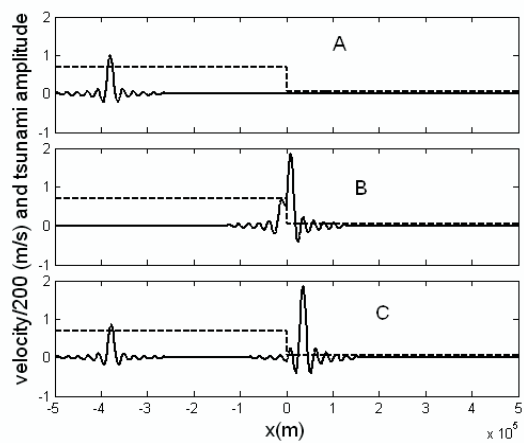,width=8cm}}
\caption[dummy0]{Evolution of a pulse propagating from a deep sea to a shallow sea (solid line). 
The dashed line is the wave velocity in units of 1/200 m/s. The upper frame (A) shows the pulse before 
the velocity discontinuity, the middle frame (B) at the velocity discontinuity and the lower 
frame (C) after the discontinuity.} \label{fig1}
\end{figure}

Every time the tsunami encounters a step, part is transmitted and part is reflected. Then, consider a 
wave with an amplitude given by $y=\cos(kx-\omega t)$, where $k$ and $\omega$ are, respectively, the wave 
number and the frequency incoming in a region where the depth of the water, and also the wave velocity, 
have a discontinuity as represented in Fig. \ref{fig1}. 
On the left-side of the discontinuity the perturbation is given by
\begin{equation}
y_1(x,t)=\cos(kx-\omega t)+R\cos(kx+\omega t+\varphi_1),
\label{left}
\end{equation}
where $R\cos(kx+\omega t+\varphi_1)$ corresponds to the reflected wave and $\varphi_1$ is a phase 
to be determined by the boundary conditions. On the right-side of the discontinuity the wave 
amplitude is given by
\begin{equation}
y_2(x,t)=T\cos(k^\prime x-\omega t+\varphi_2),
\label{right}
\end{equation}
corresponding to the transmitted wave part. The wave numbers for $x<0$ and $x>0$ are, respectively, 
\begin{equation}
k=\frac{\omega}{v}
\label{k}
\end{equation}
and
\begin{equation}
k^\prime=\frac{\omega}{v^\prime},
\label{kprime}
\end{equation}
where $v$ and $v^\prime$ are the velocities of the wavepacket at the left and right sides of the 
discontinuity.

In order to determine $R$ and $T$ we must impose the boundary conditions at $x=0$. For any instant, 
the wave should be continuous at $x=0$: $\cos\omega t+R\cos(\omega t+\varphi_1)=T\cos(-\omega t+\varphi_2)$. 
The same should happen with the flux, $f(x,t)$, given by (see eq. (\ref{flux1}))
\begin{equation}
f(x,t)=h\frac{\partial z(x,t)}{\partial t},
\end{equation}
where $z(x,t)$ is the horizontal displacement of a transversal section of water (see equation (\ref{mcons}) for 
the relation between $z$ and $y$).

Imposing the boundary conditions $y_1(0,t)=y_2(0,t)$ and $f_1(0,t)=f_2(0,t)$ 
we can deduce $\sin\varphi_1=\sin\varphi_2=0$. Then choosing $\varphi_1=\varphi_2=0$ we 
obtain
\begin{equation}
R=\frac{k^\prime-k}{k+k^\prime}=\frac{v-v^\prime}{v+v^\prime}
\label{R}
\end{equation}
and
\begin{equation}
T=\frac{2k^\prime}{k+k^\prime}=\frac{2v}{v+v^\prime}.
\label{T}
\end{equation}

It is worthwhile to mention here that other choices of $\varphi_1$ and $\varphi_2$ will change 
the signs of $R$ and $T$. However, in this case, it will also change the phases of the reflected and 
transmitted waves. Both modifications will compensate themselves and the shape of the wave will remain 
unchanged in relation of the choice $\varphi_1=\varphi_2=0$.

Reflection and transmission are very important effects in wave propagation: every time a traveling wave 
(light, water waves, pulses in strings, etc.) encounters a discontinuity in the medium where it propagates, reflection and transmission occur. 
Since there is no energy losses, energy flux is conserved. The energy of a wave is proportional to the square of 
its amplitude \cite{MaEJP05}. Thus, the energy flux is proportional to the squared amplitude times 
the wave velocity. The energy flux of the incident wave at $x=0$ is given by
\begin{equation}
\phi_{inc}=v
\end{equation}
(the amplitude of the incident wave was chosen as 1).

The reflected and transmitted energy flux are given by
\begin{equation}
\phi_{refl}=R^2v
\end{equation}
and
\begin{equation}
\phi_{trans}=T^2v^\prime,
\end{equation}
respectively.

Using eqs. (\ref{R}) and (\ref{T}) it is immediate to show that
\begin{equation}
\phi_{inc}=\phi_{refl}+\phi_{trans}.
\label{angrel}
\end{equation}
It is worthwhile to mention here that eqs. (\ref{R}), (\ref{T}) and (\ref{angrel}) are classical textbook results 
on wave at interfaces.

Fig. \ref{fig1} shows the evolution of a wavepacket in three distinct situations: before find a step, passing the 
step and after passing the step. On the left side of $x=0$ the depth is 2000 m and on the right side 10 m. 
We can note a growth of the wavepacket amplitude after passing a step due to the velocity variation, in this case 
$k^\prime/k=14$, and, then $T=1.9$.

Using energy conservation, Margaritondo deduced that the wave amplitude, when going from a sea depth $h_1$ 
to $h_2$, increases by the factor $(h_1/h_2)^{1/4}$ \cite{MaEJP05}. According to this result, the amplitude 
in our example should grow by a factor of 3.8. However, the growth observed was 1.9. The difference is due 
to the fact that when a wave packet encounters a discrete step of velocity, 
part of the energy is reflected (the reader can verify that eq. (\ref{angrel}) is satisfied). As will be 
shown in the next section, when the sea depth varies smoothly, the reflected wave can be neglected, and 
our results becomes equal to Margaritondo's result.

\section{Waves in a variable depth}

In order to study the evolution of a tsunami when it propagates in a rectangular box 
where the depth varies, we initially made a wavepacket propagating into a crescent $x$ direction. The 
variable depth was approximated by a succession of steps taken as 
narrow as we wish.

The evolution of the wave packet was calculated as follows:

\begin{itemize}
\item At time $t$, the wave packet amplitude $y(x,t)$ was divided in $n$ small discrete transversal 
sections of length  $\Delta x$ (in our case $\Delta x$ = 50000 m).
\item Every small part of the wave packet $y(x,t)$ was investigated:
\subitem (i) If in a time interval $\Delta t$ it stays in the same velocity step, then the wave packet at 
$t+\Delta t$ was simply increased by $y(x,t)$ at the position $x+v(x)\Delta t$.
\subitem (ii) If in a time interval $\Delta t$, we choose $\Delta t$ as 30 s, it encounters a velocity step, part is 
reflected and part is transmitted. The reflected and transmitted parts were calculated from eqs. (\ref{R}) and (\ref{T}) 
and added to the wave packet at $t+\Delta t$ propagating to the left or to the right, respectively. The step width and 
the time interval $\Delta t$ were chosen such that never the reflected or transmitted parts encounter a 
second step.
\end{itemize}

\begin{figure}[ht]
\centerline{\epsfig{figure=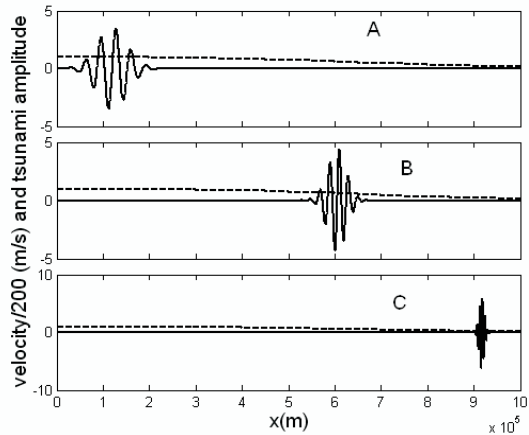,width=8cm}}
\caption[dummy0]{The upper frame (A) shows an initial wave packet traveling 
to the right. The frames (B) and (C) show, respectively, the tsunami in an intermediary position
and near to the coast. The tsunami velocity is given by the dashed line. Note that the wavepacket extension diminishes 
in the same proportion as the velocity. (Only the progressive part of the wavepacket is shown).} 
\label{fig2}
\end{figure}

Fig. \ref{fig2} shows three positions of the right-propagating wavepacket (the left-propagating was omitted).
The initial wavepacket, fig. 2A, has its center at a depth of about 3930 m ($v\sim200$ m/s), then the center of the 
wavepacket goes to a position where the depth is about 165 m ($v\sim40$ m/s), fig. 2C. The growth of the amplitude 
is about 1.7. The difference between our result and the one expected by Ref. \cite{MaEJP05} 
($(3903/165)^{0.25}=2.2$) is due to the fact that we approximate the continuous depth variation by 
discretes steps and, as a consequence, the left-propagating wavepacket was not negligible.

In the last paragraphs of this section we insert a short discussion of the refraction phenomenon. 

\begin{figure}[ht]
\centerline{\epsfig{figure=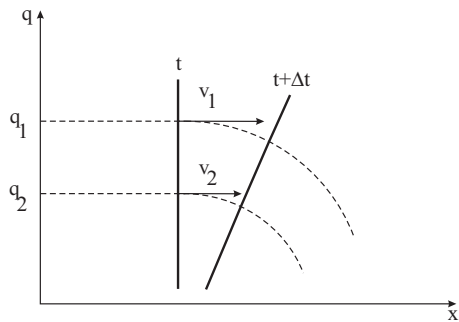,width=6cm}}
\caption[dummy0]{Change of orientation of the wave crests (solid lines) in a variable sea depth. The dashed lines 
are the trajectories of the front waves. The sea depth varies with the $q$ coordinate.}
\label{seadepth}
\end{figure}

Consider, for instance, a water wave propagating in a medium where the sea depth varies with $q$, 
as shown in fig. \ref{seadepth} (for instance $q$ can be the distance from the coast). 
The wave crest at $q_1$ has a velocity $v_1$ and at $q_2$ velocity 
$v_2$. It is a matter of geometry to show that a wavefront will change orientation as making a curve with 
radius
\begin{equation}
R=\frac{v}{\left|\frac{dv}{dq}\right|}.
\end{equation}

\begin{figure}[ht]
\centerline{\epsfig{figure=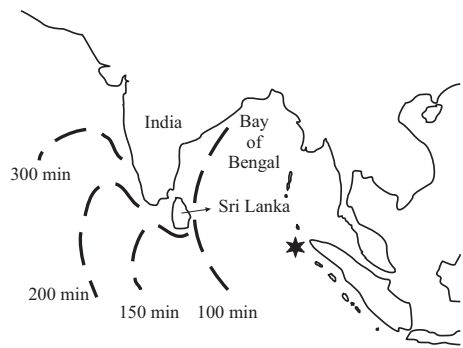,width=6cm}}
\caption[dummy0]{The dashed curves are the wavepackets corresponding, approximately, to the tsunami 
of 26 December 2004. The star shows the earthquake epicenter.}
\label{refraction}
\end{figure}

For instance, consider what happened in south of India. The seadepth varies from about 2000 m at $q_1=$500 km far from the
coast to about 100 m near the coast, $q_2\sim0$. Thus, $\left|dv/dq\right|\approx 2.2\times10^{-4}$ s$^{-1}$. As a consequence,
the radius formed by a wave crest varies from about 640 km far from the coast to about 140 km near the coast.
This is about what we can see from fig. 4.
In fig. \ref{seadepth}, a tsunami wavefront propagation, part in deep water and part in shallow water near the coast, 
will refract and, in consequence, change orientation. Fig. \ref{refraction} shows a refraction map for the December 
26, 2005 tsunami. The dashed curves are the frontwaves at 100, 150, 200 and 300 minutes after the earthquake \cite{LaSc05}.

\section{Discussion}

The tsunami of 26 December 2004, obviously, did not propagate in a rectangular box, but, approximately, 
in a circular tank. As the tsunami propagates, its extension increases and, in consequence, its amplitude 
diminishes. However, when it approaches shallow waters, near to the coast, its amplitude grew up again 
as shown by the simplified model developed in this paper. 

The developed model depends on two approximations: the wave amplitude is small and the length of 
the wavepacket is large when compared with the sea depth. Since the first approximation is not valid 
near the beach, we stopped the evolution when the front part of the tsunami wavepacket attained
a depth of about 50 m, shown in Fig. \ref{fig2}C.

In summary, with the model developed in this paper we showed how to use the approximation of shallow 
water waves to a variable depth. This simple model allows us to understand what occurs when a tsunami goes 
from deeper to shallow waters: the velocity of the rear part of the wavepacket is larger than the velocity 
of its front part, causing water to pile up. Also, refraction effects, that are not present in a sea of constant depth,
can be observed near to the coast. 

MTY thanks the Brazilian agency FAPESP (Funda\c c\~ao de Amparo a Pesquisa do Estado de S\~ao Paulo) for 
financial support.

\appendix
\section{Deduction of the wave equation for waves with wavelengths much greater than the water 
depth}

We will deduce the wave equation in a simplified 
situation. We will make the following approximations (all of them can be applied to tsunamis located 
far from the beach): the part of the restoration force that depends on the surface tension can be 
neglected in the case of waves with large wavelengths; the wavelength or the extension of the 
wavepacket will be considered much longer than the depth of the water (in the case of tsunamis 
the wavelengths and the ocean depth can have, approximately, hundreds of km and a few km, respectively); 
the wave amplitude will be considered much smaller than the ocean depth. Another 
simplification is the tank where the wave propagates: we will consider a wave 
propagating in a rectangular box with vertical walls and constant depth. In this approximation 
of shallow water waves, all the droplets in the same transversal portion have the same oscillatory horizontal 
motion along the $x$ direction. Finally, friction at the bottom will be neglected~\cite{CrAJP87}.

\begin{figure}[ht]
\centerline{\epsfig{figure=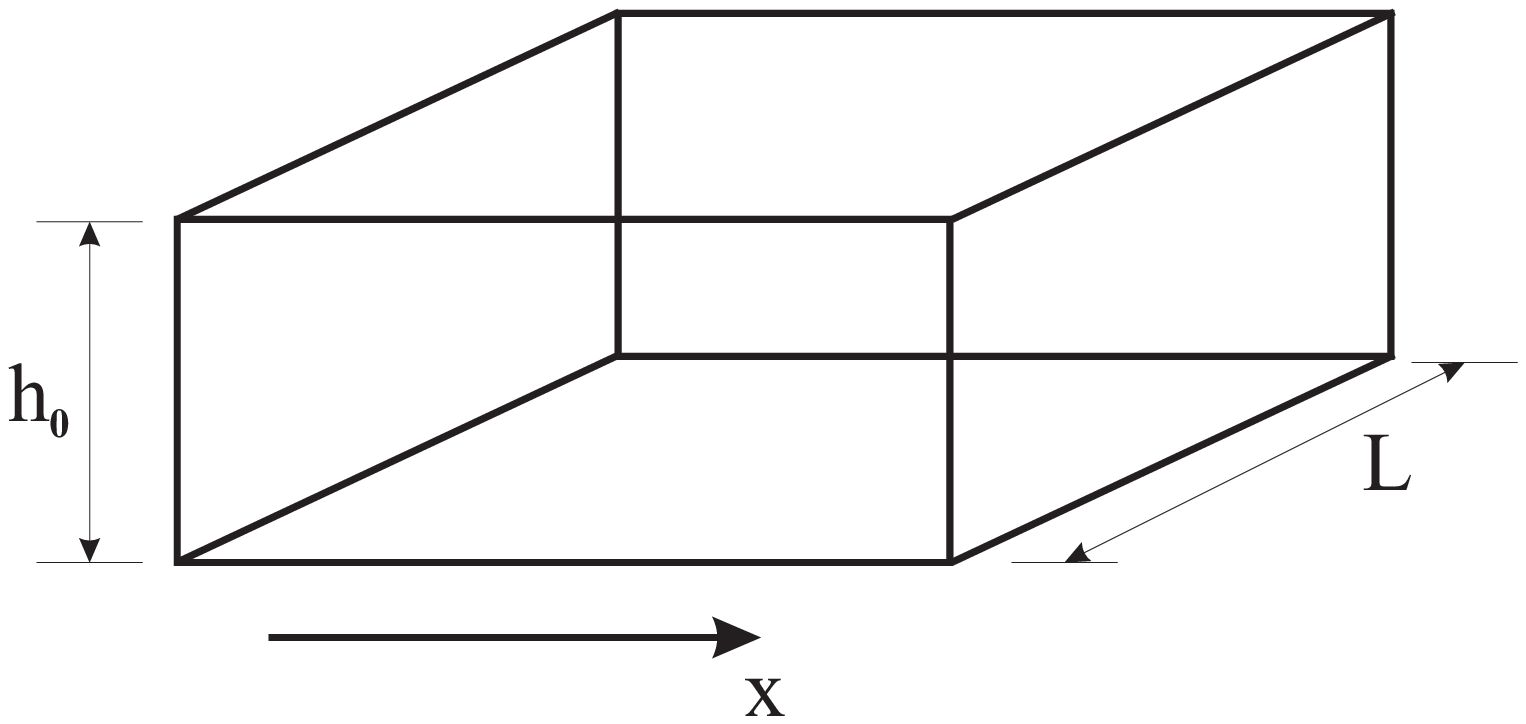,width=6cm}}
\caption[dummy0]{Rectangular box with vertical walls and constant depth where the wave 
propagates.}
\label{fig1A}
\end{figure}

Fig. \ref{fig1A} illustrates the situation considered. The wave direction of propagation is $x$; 
$h_0$ is the unperturbed height of the water; $L$ is the box width.
\begin{figure}[ht]
\centerline{\epsfig{figure=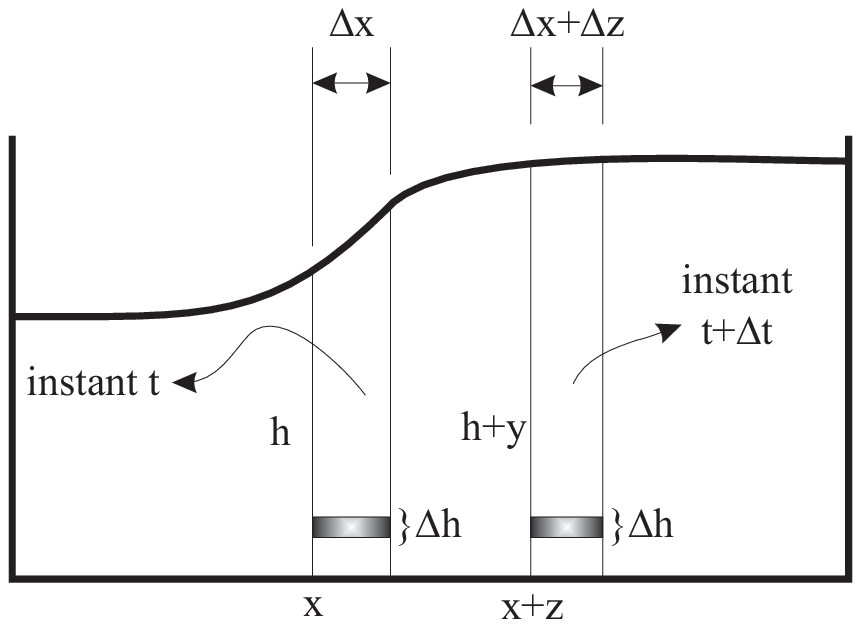,width=6cm}}
\caption[dummy0]{Side-view of Fig. \ref{fig1A} showing a perturbation in the water.}
\label{fig1B}
\end{figure}

Fig. \ref{fig1B} shows the same box of Fig. \ref{fig1A} in a side-view showing a perturbation in the water. 
A lamellar slice with width 
$\Delta x$ in $x$ and a height $h$ at a time $t$, will have a height $h+y$ and a width 
$\Delta x+\Delta z$ when it occupies the position $x+z$ at an instant $t+\Delta t$. $z=z(x,t)$ is the 
horizontal displacement - along the $x$ direction - of a vertical lamellar slice with an 
equilibrium position at $x$. When the wave propagates, this part of the water oscillates to left and right. 
Equaling the volume of water in $\Delta x$ and $\Delta x+\Delta z$, we have:
\begin{eqnarray}
\nonumber
Lh\Delta x&=&L(h+y)(\Delta x+\Delta z)\\
&=&L(h\Delta x+h\Delta z+y\Delta x+y\Delta z).
\label{flux}
\end{eqnarray}

If we consider $y<<h$ and $\Delta z<<\Delta x$, then eq. (\ref{flux}) becomes
\begin{equation}
h\Delta z+y\Delta x=0,
\end{equation}
or
\begin{equation}
y=-h\frac{\partial z}{\partial x}.
\label{mcons}
\end{equation}
This last equation is the mass conservation equation of the fluid and relates the vertical displacement of the
water surface, $y$, with the horizontal displacement of a vertical slice of water.

To apply the second Newton Law to a small portion of the lamellar slice, $\Delta h$, of water 
(see Fig. \ref{fig1B}), we should calculate the total force, $\Delta F$, acting on it. 
This force depends on the pressure difference between the opposite sides of the slice:
\begin{equation}
\Delta F=\Delta hL\left(P(x)-P(x+\Delta x)\right)\simeq-\Delta hL\frac{\partial P}{\partial x}\Delta x.
\end{equation}

Then, $F=ma$ leads to
\begin{equation}
-\Delta hL\frac{\partial P}{\partial x}\Delta x=\rho L\Delta h\Delta x\frac{\partial^2z}
{\partial t^2},
\label{force}
\end{equation}
where $m$ is the mass of the water slice, $a$ is the acceleration of the transversal section of water given 
by the second partial derivative of $z$ with respect to $t$, and $\rho$ is the water density.

Since
\begin{equation}
\frac{\partial P}{\partial x}=\rho g\frac{\partial y}{\partial x},
\label{dpress}
\end{equation}
where $g$ is the gravity acceleration, we can write eq. (\ref{force}) as
\begin{equation}
\frac{\partial^2z}{\partial t^2}=-g\frac{\partial y}{\partial x}.
\label{d2z}
\end{equation}

Derivate both sides of eq. (\ref{mcons}) with respect to $x$ we have
\begin{equation}
\frac{\partial y}{\partial x}=-h\frac{\partial^2z}{\partial x^2}.
\label{dy}
\end{equation}

Finally, using eq. (\ref{dy}) in eq. (\ref{d2z}) we obtain the wave equation~\cite{CrAJP87,alonso}
\begin{equation}
\frac{\partial^2z}{\partial t^2}=gh\frac{\partial^2z}{\partial x^2}.
\label{waveeq}
\end{equation}

Using eq. (\ref{mcons}) we can show that the vertical displacement of the water 
surface, $y$, obeys an equivalent wave equation:
\begin{equation}
\frac{\partial^2y}{\partial t^2}=gh\frac{\partial^2y}{\partial x^2}.
\label{waveeqy}
\end{equation}

The solutions of eqs. (\ref{waveeq}) and (\ref{waveeqy}) are any function of $x-vt$ or $x+vt$, 
where the wave velocity is given by
\begin{equation}
v=\sqrt{gh}.
\label{veloc}
\end{equation}

Eq. (\ref{veloc}) is a particular case of the general expression for the dispersion relation 
for waves in water surface \cite{alonso},

\begin{equation}
v=\sqrt{\left(\frac{g\lambda}{2\pi}+\frac{2\pi\sigma}{\rho\lambda}\right)\tanh\frac{2\pi h}{\lambda}},
\label{veloc2}
\end{equation}
where $\lambda$ is the wavelength, $\rho$ the water density and $\sigma$ is the water surface 
tension. In the case of long wavelength and neglecting surface tension, eq. (\ref{veloc2}) reduces to 
eq. (\ref{veloc}).

Eq. (\ref{veloc}) gives two useful conclusions for waves presenting the same characteristics 
of a tsunami: the wave velocity does not depend on the wavelength, and the wavepacket does not 
disperse when propagating in a region of constant depth.

Since $z=z(x,t)$ is the horizontal displacement of a lamellar slice of water, then the water flux, 
$f$, is given by
\begin{equation}
f=Lh\frac{\partial z}{\partial t}.
\label{flux1}
\end{equation}


\section*{References}
\begin{thebibliography}{4}
\bibitem{MaEJP05} Margaritondo G 2005 {\it Eur. J. Phys.} {\bf 26} 401
\bibitem{BeEJP01} Behroozi F and Podolefsky N 2001 {\it Eur. J. Phys.} {\bf 23} 225
\bibitem{CrAJP87} Crawford F S, 1987 {\it Am. J. Phys.} {\bf 55} 171; {\it Waves - Berkeley physics 
course volume 3} (McGraw-Hill book company, 1968).
\bibitem{LaSc05} Lay T et al., ``The great Sumatra-Andaman earthquake of 26 December 2004'',
Science {\bf 308}, 1127--1133 (2005).
\bibitem{chu} Chu A K H, {\it Eur. J. Phys.} {\bf 26} L19
\bibitem{alonso} Alonso M and Finn E J, {\it Physics} (Addison Wesley, 1992).
\end{thebibliography}
\end{document}